\documentclass[12pt]{article}
\usepackage{amsfonts}

\newcommand {\be}{\begin{equation}}
\newcommand {\ee}{\end{equation}}
\newcommand {\beg}{\begin{eqnarray}}
\newcommand {\eeg}{\end{eqnarray}}

\input{tcilatex}

\begin{document}

{}

\begin{center}
{\ Integrable hydrodynamic chains for WZNW model }\\[0pt]
\vspace{0.2cm} {Cirilo-Lombardo D.J.$^{a}$, Gershun V.D.$^{b}$} \\[0pt]
\vspace{0.2cm} {$^{a}$Joint Institute for Nuclear Research, Dubna, Russian
Federation }\\[0pt]
{$^{b}$ ITP NSC KIPT, Kharkov, Ukraine}\\[0pt]
{\footnotesize {The new integrable hydrodynamic equations obtained for WZNW
model with $SU(2)$, $SO(3)$, $SP(2)$ and $SU(\infty )$ constant torsions.\\[%
0pt]
Pacs: 02.20.Sv, 11.30.Rd, 11.40.-q, 21.60.Fw}}\\[0pt]
\textbf{{Introduction} }
\end{center}

The integrability of the two dimensional WZNW is based on the existence of
an infinite number of the local and nonlocal currents and on their charges.
The $n$-dimensional WZNW model is described by mean the chiral left $%
J_{A}^{L}=g^{-1}\partial _{A}g$ or the chiral right $J_{A}^{R}=\partial
_{A}g\;g^{-1}$ currents for arbitrary space-time dimension (${A=1,...n}$)
where $g$ is element of the group symmetry of the model. The currents $%
J_{A}=J_{A}^{\mu }t_{\mu }$ and $t_{\mu }$ are the generators of the Lie
algebra. These chiral currents were related to the left and right
multiplication on the group space. The two dimensional models ($A=0,1)$ have
the following additional chiral currents 
\[
J_{\mu }^{L}(t,x)=\frac{J_{0\mu }+\delta _{\mu \nu }J_{1}^{\nu }}{\sqrt{2}}%
=U_{\mu }(x+t),\;\;J_{\mu }^{R}(t,x)=\frac{J_{0\mu }-\delta _{\mu \nu
}J_{1}^{\nu }}{\sqrt{2}}=V_{\mu }(x-t), 
\]%
related to the dynamics on the $(t,x)$ plane. The chiral currents $U_{\mu }$%
, $V_{\mu }$ play an important role for the construction and investigation
of this type of integrable systems. We can' t separate the movement on the
left-moving mode and on the right-moving mode for the $\sigma $-model under
consideration in order to formulate the movement on only one mode. It was
did by the introduction of the Witten term to the Wess-Zumino model. This
term introduces a potential for the torsion tensor on the curved space of
the group parameters in the addition to the metric tensor. It is possible to
extract the movement on one mode with the fulfilling of some conditions
between the constant torsion tensor and the structure constant of Lie
algebra. In this work, Lagrangian and equations of motion in the repere
formalism are considered, being the antisymmetric field $B_{ab}$ obtained in
terms of the repere. Also, the Hamiltonian formalism and the commutations
relations are re-written in new variables. These variables are precisely the
chiral currents under the condition that the external torsion coincides
(anti-coincides) with the structure constants of the $SU(2)$, $SO(3)$, $%
SP(2) $ algebras. In this manner, the equation of motion for the density of
the first Casimir operator is obtained as the inviscid Burgers equation,
being its solution expressed as the Lambert function. The integrable
infinite dimensional hydrodynamic chains are constructed for WZNW model with
the constant $SU(2)$, $SO(3)$, $SP(2)$ torsions and for this model with the $%
SU(\infty )$, $SO(\infty )$, $SP(\infty )$ constant torsions. Finally, the
new equations of motion of hydrodynamic type are explicitly obtained for the
initial chiral currents in terms of the symmetric structure constant of the $%
SU(\infty )$, $SO(\infty )$, $SP(\infty )$ algebras.

\begin{center}
\textbf{{Lagrangian and equation of motion} }
\end{center}

The conformal invariant two-dimensional non-linear sigma model is described
by WZNW model which is the sigma model [1--4] with Wess-Zumino term [5--8]
on the group manifold. To each point of a 2-dimensional world-sheet one
associate an element $g$ of a group $G$. We want to construct an action with
the Lagrangian density which is the element of volume of the two-dimensional
space invariant under the group transformations 
\begin{equation}
S=\frac{1}{4}\int \frac{Tr(\omega \wedge dx^{\alpha })(\omega \wedge
dx^{\beta })\eta _{\alpha \beta }}{\epsilon _{\lambda \rho }dx^{\lambda
}\wedge dx^{\rho }}+\frac{1}{2}\int Tr{(\omega (d)\wedge \omega (d)\wedge
\omega (d))}.  \label{eq1}
\end{equation}%
Here $x^{\alpha }=(t,x)$ are coordinates of the flat two-dimensional space: $%
\alpha =(0,1)$ with signature $(-1,1)$ and $\eta _{\alpha \beta }$ is the
diagonal metric of this space. The form $\omega (d)=\omega (d)^{\mu }t_{\mu
} $ is the differential Cartan one-form which belongs to a simple Lie
algebra 
\begin{equation}
\lbrack t_{\mu },t_{\nu }]=2iC_{\mu \nu }^{\lambda }t_{\lambda },\;Tr(t_{\mu
}t_{\nu })=2g_{\mu \nu },\;(\mu ,\nu =1,2,...,n)\;.  \label{eq2}
\end{equation}%
In any parametrization, Cartan forms $\omega (d)=(g^{-1}dg)^{\mu }t_{\mu }$
depend on the group parameters $\phi ^{a}$: $\omega (d)=\omega (\phi ,d\phi
).$ The first term of the Lagrangian (\ref{eq1}) has form 
\begin{equation}
\frac{Tr(\omega \wedge dx^{\alpha })(\omega \wedge dx^{\beta })\eta _{\alpha
\beta }}{\epsilon _{\lambda \rho }dx^{\lambda }\wedge dx^{\rho }}%
=2g_{ab}(\phi )\frac{\partial \phi ^{a}}{\partial x^{\alpha }}\frac{\partial
\phi ^{b}}{\partial x^{\beta }}\eta ^{\alpha \beta }d^{2}x.  \label{eq3}
\end{equation}%
Here we introduce the notation 
\begin{equation}
g_{ab}(\phi )=g_{\mu \nu }\omega _{a}^{\mu }\omega _{b}^{\nu },\;dx^{\gamma
}\wedge dx^{\alpha }=\epsilon ^{\gamma \alpha }d^{2}x.  \label{eq4}
\end{equation}%
One can see, that $g_{ab}(\phi )$ is metric tensor on the curved space of
local fields $\phi ^{a},\;(a=1,2,...,n)\;.$The $\omega _{a}^{\mu }(\phi )$
forms a repere basis on the tangent space with the metric $g_{\mu \nu }$ in
arbitrary point of the curved space $\phi ^{a}$. Here we want to rewrite a
WZNW model as sigma - model of string type,\ equipped with an antisymmetric
field $B_{ab}(\phi ),$ in terms of the repere for the arbitrary metric $%
g_{ab}(\phi )$ and for any dimension $n$: 
\begin{equation}
Tr{(\omega (d)\wedge \omega (d)\wedge \omega (d))}=g_{\mu \nu }(\Omega
_{a}^{\mu }\frac{\partial \Omega _{b}^{\nu }}{\partial x^{C}}-\Omega
_{b}^{\mu }\frac{\partial \Omega _{a}^{\nu }}{\partial x^{C}})\frac{\partial
\Phi ^{a}}{\partial x^{A}}\frac{\partial \Phi ^{b}}{\partial x^{B}}\epsilon
^{ABC}d^{3}x.  \label{eq5}
\end{equation}%
The integrability condition: $\partial _{A}g=g\Omega _{A}^{\mu }t_{\mu }$
was used 
\[
\partial _{A}\Omega _{B}^{\mu }-\partial _{B}\Omega _{A}^{\mu }+2iC^{\mu \nu
\lambda }\Omega _{A}^{\nu }\Omega _{B}^{\lambda }=0. 
\]%
Here $x^{A}$ $(A=0,1,2)$ are coordinates of the three dimensional
space-time, $\Omega ^{\mu }(d)$ is a one form on this space. Let us to
separate the last component of index $A$ $(A=\alpha ,2;\;\alpha =0,1)$ in
the equation (\ref{eq5}). Then, the second term of the action has the
following form: 
\begin{equation}
\int g_{\mu \nu }\epsilon ^{\alpha \beta 2}(\Omega _{a}^{\mu }\partial
_{2}\Omega _{b}^{\nu }-\partial _{2}\Omega _{a}^{\mu }\Omega _{b}^{\nu })%
\frac{\partial \Phi ^{a}}{\partial x^{\alpha }}\frac{\partial \Phi ^{b}}{%
\partial x^{\beta }}d^{3}x=\int d^{2}x\int\limits_{0}^{M}\epsilon ^{\alpha
\beta 2}B_{ab2}\frac{\partial \Phi ^{a}}{\partial x^{\alpha }}\frac{\partial
\Phi ^{b}}{\partial x^{\beta }}dx^{2}  \label{eq6}
\end{equation}%
Here $B_{ab2}=g_{\mu \nu }(\Omega _{a}^{\mu }\partial _{2}\Omega _{b}^{\nu
}-\partial _{2}\Omega _{a}^{\mu }\Omega _{b}^{\nu })=-B_{ba2}.$ We will
integrate on the coordinate $x^{2}$ in the limits ( $0$ , $M$ ) with the
following boundary conditions: 
\[
\Phi ^{a}(x^{\alpha },x^{2})\left. {}\right\vert ^{x^{2}=M}=\phi
^{a}(x^{\alpha }),B_{ab2}(x^{\alpha },x^{2})\left. {}\right\vert
^{x^{2}=M}=B_{ab}(x^{\alpha }). 
\]%
The integral in $x^{2}$ on the lower limit of integration equals zero, what
is easily seen by using the expansion of the integrand into the Taylor
series. Consequently, the total action is: 
\begin{equation}
S=\frac{1}{2}\int d^{2}x[g_{ab}(\phi )\eta ^{\alpha \beta }+B_{ab}(\phi
)\epsilon ^{\alpha \beta }]\frac{\partial {\phi ^{a}}}{\partial {x^{\alpha }}%
}\frac{\partial {\phi ^{b}}}{\partial {x^{\beta }}}.  \label{eq7}
\end{equation}%
Here $g_{ab}(\phi )=g_{ba}(\phi )$ is the metric tensor of the group space $%
G $ and $\phi ^{a}(x)$ are the group parameters, $a,b=1,2,...n$. The
background field $B_{ab}(\phi )$ on the group space $G$ is the antisymmetric
tensor field $B_{ab}(\phi (x))=-B_{ba}(\phi (x))$. The coordinates $%
x^{\alpha }=(t,x),\;\alpha =0,1$ belong to the 2-dimensional word-sheet with
the constant metric tensor $\eta _{\alpha \beta }$ and the signature $(-1,1)$%
. Let us introduce a repere $e_{\mu }^{a}(\phi )=\omega _{\mu }^{a}$ on the
compact group space $G$ and its inverse $e_{a}^{\mu }(\phi )=\omega
_{a}^{\mu }$ such that the metric tensor can be explicitly written as 
\begin{equation}
g_{ab}(\phi )=e_{a}^{\mu }(\phi )e_{b}^{\nu }(\phi )\delta _{\mu \nu
},\;\delta _{\mu \nu }=e_{\mu }^{a}(\phi )e_{\nu }^{b}(\phi )g_{ab}(\phi ).
\label{eq8}
\end{equation}%
Here $\delta _{\mu \nu }$ $(\mu ,\nu =1,2,...n)$ is a constant tensor on the
tangent space of the compact group space $G$ at some point $\phi ^{a}(x)$
with the same signature as $g_{ab}(\phi )$. To introduce the Hamiltonian, we
rewrite the Lagrangian density and the equation of motion in terms the
world-sheet coordinates $(t,x)$ 
\begin{equation}
L=\frac{1}{2}g_{ab}(\phi )[\frac{\partial \phi ^{a}}{\partial t}\frac{%
\partial \phi ^{b}}{\partial t}-\frac{\partial \phi ^{a}}{\partial x}\frac{%
\partial \phi ^{b}}{\partial x}]+B_{ab}(\phi )\frac{\partial \phi ^{a}}{%
\partial t}\frac{\partial \phi ^{b}}{\partial x}.  \label{eq9}
\end{equation}%
Then, the equation of motion takes the form: 
\begin{equation}
g_{ab}(\phi )[\frac{\partial ^{2}\phi ^{a}}{{\partial t}{\partial t}}-\frac{%
\partial ^{2}\phi ^{a}}{{\partial x}{\partial x}}]+\Gamma _{abc}(\phi )[%
\frac{\partial \phi ^{b}}{\partial t}\frac{\partial \phi ^{c}}{\partial t}-%
\frac{\partial \phi ^{b}}{\partial x}\frac{\partial \phi ^{c}}{\partial x}%
]+2H_{abc}(\phi )\frac{\partial \phi ^{b}}{\partial t}\frac{\partial \phi
^{c}}{\partial x}=0.  \label{eq10}
\end{equation}%
\begin{equation}
\Gamma _{abc}=\frac{1}{2}(\frac{\partial g_{ab}}{\partial \phi ^{c}}+\frac{%
\partial g_{ac}}{\partial \phi ^{b}}-\frac{\partial g_{bc}}{\partial \phi
^{a}}),H_{abc}=\frac{\partial B_{ab}}{\partial \phi ^{c}}+\frac{\partial
B_{ca}}{\partial \phi ^{b}}+\frac{\partial B_{bc}}{\partial \phi ^{a}}.
\label{eq11}
\end{equation}%
where $\Gamma _{abc}(\phi )$ are the Christoffel symbols. It is a symmetric
function in $b,c$. The canonical momentum is as follows 
\begin{equation}
p_{a}(\phi (t,x))=\frac{\delta L}{\delta (\frac{\partial \phi ^{a}}{\partial
t})}=g_{ab}(\phi )\frac{\partial \phi ^{b}}{\partial t}+B_{ab}(\phi )\frac{%
\partial \phi ^{b}}{\partial x}.  \label{eq12}
\end{equation}%
By definition, the Hamiltonian is 
\begin{equation}
H(\phi ,p)=\frac{1}{2}g^{ab}(\phi )[p_{a}-B_{ac}(\phi )\frac{\partial \phi
^{c}}{\partial x}][p_{b}-B_{bd}(\phi )\frac{\partial \phi ^{d}}{\partial x}]+%
\frac{1}{2}g_{ab}(\phi )\frac{\partial \phi ^{a}}{\partial x}\frac{\partial
\phi ^{b}}{\partial x}.  \label{eq13}
\end{equation}%
Now, let us introduce new dynamical variables as follows: 
\begin{equation}
J_{0\mu }(\phi )=e_{\mu }^{a}(\phi )[p_{a}-B_{ab}(\phi )\frac{\partial \phi
^{b}}{\partial x}],\;J_{1\mu }(\phi )=e_{a}^{\mu }(\phi )\frac{\partial \phi
^{a}}{\partial x}.  \label{eq14}
\end{equation}%
We see that the Hamiltonian (\ref{eq13}) is factorized in these variables 
\begin{equation}
H=\frac{1}{2}[\delta ^{\mu \nu }J_{0\mu }(\phi )J_{0\nu }(\phi )+\delta
_{\mu \nu }J_{1}^{\mu }(\phi )J_{1}^{\nu }(\phi )].  \label{eq15}
\end{equation}%
The equations of motion in terms of this variables are of first order: 
\begin{equation}
\partial _{0}J_{1}^{\mu }(\phi )-\partial _{1}J_{0}^{\mu }(\phi )=C_{\nu
\lambda }^{\mu }J_{0}^{\nu }(\phi )J_{1}^{\lambda }(\phi ),\;\;\partial
_{0}J_{0}^{\mu }(\phi )-\partial _{1}J_{1}^{\mu }(\phi )=-H_{\nu \lambda
}^{\mu }(\phi )J_{0}^{\nu }(\phi )J_{1}^{\lambda }(\phi ).  \label{eq16}
\end{equation}%
Here $C^{\mu \nu \lambda }$ is the structure constant tensor which can be
obtained from the Maurer-Cartan equation: 
\begin{equation}
C_{\nu \lambda }^{\mu }=\frac{\partial e_{a}^{\mu }(\phi )}{\partial x^{b}}%
[e_{\nu }^{b}(\phi )e_{\lambda }^{a}(\phi )-e_{\nu }^{a}(\phi )e_{\lambda
}^{b}(\phi )]=[\frac{\partial e_{a}^{\mu }(\phi )}{\partial x^{b}}-\frac{%
\partial e_{b}^{\mu }(\phi )}{\partial x^{a}}]e_{\nu }^{b}(\phi )e_{\lambda
}^{a}(\phi )  \label{eq17}
\end{equation}%
and the canonical Poisson bracket (PB) is: 
\begin{equation}
\{\phi ^{a}(x),p_{b}(y)\}=\delta _{b}^{a}\delta (x-y).  \label{eq18}
\end{equation}%
Now, we consider the commutation relations for the functions $J_{0\mu }(\phi
(x)),\;J_{1\mu }(\phi (x))=\delta _{\mu \nu }J_{1}^{\nu }(\phi (x))$ on the
phase space under the PB (\ref{eq18}) 
\[
\{J_{0\mu }(\phi (x)),\;J_{0\nu }(\phi (y))\}=C_{\mu \nu }^{\lambda
}J_{0\lambda }(\phi (x))\delta (x-y)+H_{\mu \nu }^{\lambda }(\phi
(x))J_{1\lambda }(\phi (x))\delta (x-y), 
\]%
\[
\{J_{0\mu }(\phi (x)),\;J_{1\nu }(\phi (y))\}=C_{\mu \nu }^{\lambda
}J_{1\lambda }(\phi (x))\delta (x-y)+g_{\mu \nu }\frac{\partial }{\partial x}%
\delta (x-y), 
\]%
\begin{equation}
\{J_{1\mu }(\phi (x)),\;J_{1\nu }(\phi (y))\}=0.  \label{eq19}
\end{equation}%
Let us introduce the chiral variables 
\begin{equation}
U_{\mu }=\frac{J_{0\mu }+\delta _{\mu \nu }J_{1}^{\nu }}{\sqrt{2}}%
,\;\;V_{\mu }=\frac{J_{0\mu }-\delta _{\mu \nu }J_{1}^{\nu }}{\sqrt{2}}.
\label{eq20}
\end{equation}%
The commutation relations for the chiral currents $U^{\mu }(\phi ),V^{\mu
}(\phi )$ are not Poisson brackets because the torsion $H_{\mu \nu
}^{\lambda }(\phi )$ is not a smooth function. These commutation relations
form an algebra, if $H_{\mu \nu }^{\lambda }(\phi )$ is a constant tensor.
The interesting cases arise if $H_{\mu \nu }^{\lambda }=\pm C_{\mu \nu
}^{\lambda }.$ In the case $H_{\mu \nu }^{\lambda }=-C_{\mu \nu }^{\lambda }$
the variables $U_{\mu }(\phi )$ form the closed Kac-Moody algebra [9, 10]
for the right chiral currents: 
\begin{equation}
\{U_{\mu }(\phi (x)),\;U_{\nu }(\phi (y))\}_{2}=C_{\mu \nu }^{\lambda
}U_{\lambda }(\phi (x))\delta (x-y)+\delta _{\mu \nu }\partial _{x}\delta
(x-y).  \label{eq21}
\end{equation}%
Here we have been noted the PB (\ref{eq21}) as $PB_{2}$. The last relations
are not essential. In the case of $H_{\mu \nu }^{\lambda }=C_{\mu \nu
}^{\lambda },$ the variables $V_{\mu }(\phi )$ form the closed Kac-Moody
algebra for the left chiral currents 
\begin{equation}
\{V_{\mu }(\phi (x)),\;V_{\nu }(\phi (y))\}=C_{\mu \nu }^{\lambda
}V_{\lambda }(\phi (x))-\delta _{\mu \nu }\partial _{x}\delta (x-y).
\label{eq22}
\end{equation}%
Notice that the Kac-Moody algebra [9, 10] has been considered as a hidden
symmetry of the two-dimensional chiral models [11]. In the 1983 one of the
authors (VDG) with Volkov and Tkach [12] considered the algebra of the
nonlocal charges in $\sigma $-model in the frame of the integrability of
this model. We shown in this previous reference that the nonlocal charges
form the enveloped algebra over the Kac-Moody algebra. If $C_{\mu \nu
}^{\lambda }=H_{\mu \nu }^{\lambda }$ the equation of motion is 
\begin{equation}
\partial _{+}V_{\mu }(\phi (t,x))=0,\;\;\partial _{-}U_{\mu }(\phi
(t,x))=C_{\mu }^{\nu \lambda }V_{\nu }(\phi )U_{\lambda }(\phi ).
\label{eq23}
\end{equation}%
We see from the equations (\ref{eq21}) and (\ref{eq23}) that the chiral
currents $U_{\mu }$ form the closed system in the first case and, from the
equations (\ref{eq22}), (\ref{eq23}), that the chiral currents $V_{\mu }$
also form the closed system in the second case. Precisely, the chiral
currents are the generators of group transformations with the structure
constants $C_{\lambda }^{\mu \nu }$ in the tangent space.

\begin{center}
\textbf{Integrable WZNW model with constant torsion}
\end{center}

The components of the torsion $C_{abc}$ are the structure constants of the
Lie algebra. In the bi-Hamiltonian approach to the integrable string models
with the constant torsion, we have considered the conserved primitive chiral
invariant currents (densities of the dynamical Casimir operators) $%
C_{n}(U(x))$, as the local fields of a Riemmann manifold [13,14]. The
primitive and non-primitive local charges of the invariant chiral currents
form the hierarchy of the new Hamiltonians. The primitive invariant currents
are the densities of the Casimir operators, in contrast, the non primitive
currents are functions of the primitive ones. The commutation relations (\ref%
{eq21}) show that the currents $U^{\mu }$ form the closed algebra.
Therefore, we will consider PBs of the right chiral currents $U^{\mu }$ and
the Hamiltonians constructed only from the right currents. The constant
torsion does not contributes to the equations of motion, but it gives the
possibility to introduce the group structure and the symmetric structure
constants. This paper was stimulated by the papers [16, 17] concerning the
local conserved charges in two dimensional models. In [16] the local
invariant chiral currents, as polynomials of the initial chiral currents of
the $SU(n)$, $SO(n)$, $SP(n),$ were constructed for principal chiral models.
Their paper [16] was based on the [17] involving the invariant tensors for
the simple Lie algebras. Let us take $t_{\mu }$ the generators of the $SU(n)$%
, $SO(n)$, $SP(n)$ Lie algebras (\ref{eq2}). There are additional relations
for the generators of the Lie algebra in the defining matrix representation.
There is the following relation for the symmetric double product of the
generators of $SU(n)$ algebra: 
\begin{equation}
\{t_{\mu },t_{\nu }\}=\frac{4}{n}\delta _{\mu \nu }+2d_{\mu \nu \lambda
}t_{\lambda },\,\,\mu =1,...,n^{2}-1.  \label{eq24}
\end{equation}%
where $d_{\mu \nu \lambda }$ is a totally symmetric structure constant
tensor. The Killing tensor $g_{\mu \nu }$ equals $\delta _{\mu \nu }$ for
the compact Lie algebras. Similar relation for the totally symmetric triple
product of the $SO(n)$ and $SP(n)$ algebras has the form: 
\begin{equation}
t_{(\mu }t_{\nu }t_{\lambda )}=v_{\mu \nu \lambda }^{\rho }\,t_{\rho }.
\label{eq25}
\end{equation}%
where $v_{\mu \nu \lambda \rho }$ is a totally symmetric structure constant
tensor. The invariant chiral currents are the Liouville coordinates and they
can be constructed as the product of the invariant symmetric tensors: 
\[
d_{(\mu _{1}...\mu _{n})}=d_{({\mu _{1}\mu _{2}}}^{k_{1}}d_{\mu
_{3}k_{1}}^{k_{2}}...d_{\mu _{n-1}\mu _{n})}^{k_{n-3}},\;d_{\mu _{1}\mu
_{2}}=\delta _{\mu _{1}\mu _{2}} 
\]%
For the $SU(n)$ group and the initial chiral currents $U^{\mu }(\phi (x))$
we have: 
\begin{equation}
C_{n}(U(\phi (x)))=d_{(\mu _{1}...\mu _{n})}U_{\mu _{1}}U_{\mu
_{2}}...U_{\mu _{n}},\;C_{2}(U(\phi (x)))=\delta _{\mu \nu }U^{\mu }U^{\nu }.
\label{eq26}
\end{equation}%
Analogically, similar construction can be used for $SO(n)$, $SP(n)$ groups.
The invariant chiral currents can be constructed as product of the invariant
symmetric constant tensors: 
\[
v_{(\mu _{1}...\mu _{2n})}=v_{(\mu _{1}\mu _{2}\mu _{3}}^{\nu _{1}}v_{\mu
_{4}\mu _{5}}^{\nu _{1}\nu _{2}}...v_{\mu _{2n-2}\mu _{2n-1}\mu _{2n})}^{\nu
_{2n-3}},\;v_{\mu _{1}\mu _{2}}=\delta _{\mu _{1}\mu _{2}}. 
\]%
and the corresponding initial chiral currents $U^{\mu }$: 
\begin{equation}
C_{2n}(U(\phi (x)))=v_{\mu _{1}...\mu _{2n}}U^{\mu _{1}}...U^{\mu
_{2n}},\;C_{2}(U(\phi (x)))=\delta _{\mu _{1}\mu _{2}}U^{\mu _{1}}U^{\mu
_{2}}.  \label{eq27}
\end{equation}%
The invariant chiral currents for $SU(2)$, $SO(3)$, $SP(2)$ have the form: 
\begin{equation}
C_{2n}=(C_{2})^{n}  \label{eq28}
\end{equation}%
Another family of the invariant symmetric currents $J_{n}$ based on the
invariant symmetric chiral currents of simple Lie groups, are realized as
the symmetric trace of the $n$ product chiral currents $U(x)=t_{\mu }U^{\mu
},$ $\mu =1,...,n^{2}-1$ 
\begin{equation}
J_{n}(U(\phi (x)))=SymTr(U...U).  \label{eq29}
\end{equation}%
These invariant currents are the polynomials of the product of the basic
chiral currents $C_{k},\;k=2,3,...,k$ [13, 14]. Let us introduce the PB of
hydrodynamic type for the chiral currents in the Liouvlle form [18]: 
\begin{equation}
\{C_{m}(\phi )(x),C_{n}(\phi (y))\}=-W_{mn}(\phi (y))\frac{\partial }{%
\partial y}\delta (y-x)+W_{nm}(\phi (x))\frac{\partial }{\partial x}\delta
(x-y).  \label{eq30}
\end{equation}%
The asymmetric Hamiltonian function $W_{mn}(U(\phi (x)))$ for the finite
dimensional $SU(n)$, $SO(n)$, $SP(n)$ group has the following form: 
\begin{equation}
W_{mn}(C(U(x)))=\frac{n-1}{m+n-2}\sum_{k}a_{k}C_{m+n-2,k}(U(x)),\;%
\sum_{k=0}a_{k}=mn.  \label{eq32}
\end{equation}%
This PB can be rewritten as the PB of the hydrodynamic type by use  the
following equalities: 
\[
B(y)A(x)\frac{\partial }{\partial x}\delta (x-y)=B(y)A(y)\frac{\partial }{%
\partial x}\delta (x-y)-B(y)\frac{\partial A(y)}{\partial y}\delta (x-y), 
\]%
\[
\frac{\partial A(y)}{\partial y}\delta (x-y)+A(x)\frac{\partial }{\partial x}%
\delta (x-y)=A(y)\frac{\partial }{\partial x}\delta (x-y),\;\;\frac{\partial 
}{\partial x}\delta (x-y)=-\frac{\partial }{\partial y}\delta (y-x). 
\]%
Above, the invariant total symmetric currents $C_{n,k},\;k=1,2...$ are new
currents, polynomials of the product of the basic invariant currents $%
C_{n_{1}}C_{n_{2}}...C_{n_{n}}$, $n_{1}+...+n_{n}=n$. They can be obtained
by mean the explicit computation of the total symmetric invariant currents $%
J_{n}$ using the different replacements of the double product (\ref{eq24})
for the $SU(n)$ group and of the triple product (\ref{eq25}) for the $SO(n)$%
, $SP(n)$ groups, into the expressions for the invariant currents $J_{n}$
[13]. Here are only $l=n-1$ primitive invariant tensors for $SU(n)$ algebra, 
$l=\frac{n-1}{2}$ for $SO(n)$ algebra and $l=\frac{n}{2}$ for $SP(n)$
algebra. Higher invariant currents $C_{n}$ for $n\geq l+1$ are non-primitive
currents and they are polynomials of primitive currents. By using formula (%
\ref{eq30}) we can obtain the expression for these polynomials within the
condition $J_{k}=0$ for $k>l$ for the generating function:\noindent 
\[
det(1-\lambda t_{\mu }U^{\mu })=\exp {Tr(ln(1-\lambda U))}=\exp {%
(-\sum_{k=2}^{\infty }\frac{\lambda ^{k}}{k}J_{k})}. 
\]%
The corresponding charges for non-primitive chiral currents $C_{n}$ are not
Casimir operators. Consequently the WZNW model is not an integrable system
for the group symmetry of the finite rank $l\geq 1$. \noindent

\begin{center}
\textbf{Integrable WZNW models with $SU(2)$, $SO(3)$, $SP(2)$ constant
torsions}
\end{center}

There is one primitive invariant tensor for the algebras of $SU(2)$, $SO(3)$%
, $SP(2)$. As we have been pointed out, the invariant non primitive tensors
for $n\geq 2$ are functions of the primitive tensors. Let us to introduce
the local chiral currents based on the invariant symmetric polynomials on
the $SU(2)$, $SO(3)$, $SP(2)$ Lie groups: 
\[
C_{2}(U)=\delta _{\mu \nu }U^{\mu }U^{\nu },C_{2n}(U)=(\delta _{\mu \nu
}U^{\mu }U^{\nu })^{n}, 
\]%
where $n=1,2,...$ and $\mu ,\nu =1,2,3.$ The PB of Liouville coordinate $%
C_{2}(U(x))$ has the following form: 
\[
\{C_{2}(U(x)),C_{2}(U(y))\}=-2C_{2}(U(y))\partial _{y}\delta
(y-x)+2C_{2}(U(x))\partial _{x}\delta (x-y), 
\]%
We will consider the invariant chiral $C_{2}(U(x))$ as a local field on the
Riemmann space of the chiral currents. As the Hamiltonians we choose the
following functions: 
\begin{equation}
H_{2(n+1)}=\frac{1}{2(n+1)}\int\limits_{0}^{2\pi
}C_{2}^{n+1}(U(y))dy,\;n=0,1,...\infty .  \label{eq32}
\end{equation}%
The equation of motion for the density of the first Casimir operator is as
follows: 
\begin{equation}
\frac{\partial C_{2}}{\partial t_{2(n+1)}}-(2n+1)(C_{2})^{n}\frac{dC_{2}}{dx}%
=0.  \label{eq33}
\end{equation}%
and the equation for the currents $C_{2}^{n}=C_{2n}$ is: 
\begin{equation}
\frac{\partial C_{2}^{n}}{\partial \tau _{n}}+(C_{2})^{n}\frac{dC_{2}^{n}}{dx%
}=0,\;\tau _{n}=-(2n+1)t_{2(n+1)}.  \label{eq34}
\end{equation}%
The above equation is precisely inviscid Burgers equation. We will find the
solution in the form: 
\begin{equation}
C_{2}^{n}(\tau _{n},x)=\exp {(a+i(x-\tau _{n}C_{2}^{n}(\tau _{n},x)))}.
\label{eq35}
\end{equation}%
To obtain the solution of equation (\ref{eq34}) is convenient to rewrite
this equation of motion as: 
\begin{equation}
Y_{n}=Z_{n}e^{Z_{n}},\;Y_{n}=i\tau _{n}e^{(a+ix)},\;Z_{n}=i\tau
_{n}C_{2}^{n}.  \label{eq36}
\end{equation}%
Then, the inverse transformation $Z_{n}=Z_{n}(Y_{n})$ is defined by mean the
periodical Lambert function [14]: 
\begin{equation}
C_{2}^{n}(\tau _{n},x)=\frac{1}{i\tau _{n}}W(i\tau _{n}e^{a+ix}).
\label{eq37}
\end{equation}%
Consequently, the solution for the first Casimir operator is: 
\begin{equation}
C_{2}(t_{2(n+1)},x)=[\frac{i}{(2n+1)t_{2(n+1)}}%
W(-i(2n+1)t_{2(n+1)}e^{a+ix})]^{\frac{1}{n}}.  \label{eq38}
\end{equation}%
With these results, the equation of motion for the initial chiral current $%
U^{\mu }$ defined by the PB (\ref{eq21}) and the Hamiltonian (\ref{eq32})
is: 
\begin{equation}
\frac{\partial U_{\mu }}{\partial t_{2(n+1)}}=\frac{\partial }{\partial x}%
[U_{\mu }(UU)^{n}]=nU_{\mu }C_{2}^{n-1}\frac{\partial }{\partial x}%
C_{2}+C_{2}^{n}\frac{\partial }{\partial x}U_{\mu },\;\mu =1,2,3.
\label{eq39}
\end{equation}%
It is easy to test, that equation of motion (\ref{eq33}) is in fully
agreement with equation (\ref{eq39}) simply by multiplication with the
chiral current $U_{\mu }$ on the both sides of equation (\ref{eq39}). It is
possible to rewrite this equation as a linear equation by using the solution
(\ref{eq37}) which diagonalize the equation (\ref{eq39}): 
\[
\frac{\partial U_{\mu }}{\partial t_{2(n+1)}}=\frac{\partial U_{\mu }}{%
\partial x}f_{n}+U_{\mu }\frac{\partial }{\partial x}f_{n} 
\]%
or as the linear nonhomogeneous equation: 
\begin{equation}
\frac{\partial z^{\mu }}{\partial t_{2(n+1)}}=f_{n}(t_{n},x)\frac{\partial
z^{\mu }}{\partial x}+\frac{\partial }{\partial x}f_{n}(t_{n},x),\;\;z^{\mu
}=\ln U^{\mu },\;f_{n}=C_{2}^{n},\;\frac{\partial z^{\mu }}{\partial x}=%
\frac{1}{U^{\mu }}\frac{\partial U^{\mu }}{\partial x},\;(not\;sum).
\label{eq40}
\end{equation}

\begin{center}
\textbf{Infinite dimensional hydrodynamic chains}
\end{center}

The first example of the infinite dimensional hydrodynamic chains is based
on the invariant chiral currents $C_{2n}=(C_{2})^{n},\;n=1,2,...,\infty $ of
the WZNW model with the $SU(2)$, $SO(3)$, $SP(2)$ constant torsions. The PB
of the different degrees of the invariant chiral currents $C_{2}^{n}(x)$, $%
C_{2}^{m}(x)$ has form: 
\begin{equation}
\{C_{2}^{m}(x),C_{2}^{n}(y)\}=\frac{2nm(2m-1)}{n+m-1}C_{2}^{n+m-1}(x)\frac{%
\partial \delta (x-y)}{\partial x}-\frac{2nm(2n-1)}{n+m-1}C_{2}^{n+m-1}(y)%
\frac{\partial \delta (y-x)}{\partial y}.  \label{eq41}
\end{equation}%
The equation of motion for invariant current $C_{2}^{m}$ with Hamiltonian 
\[
H_{2n}=\frac{1}{2n}\int\limits_{0}^{2pi}C_{2n}(y)dy 
\]%
has the form: 
\[
\frac{\partial C_{2}^{m}}{\partial t_{2n}}=\frac{m(2n-1)}{m+n-1}\frac{%
\partial C_{2}^{m+n-1}}{\partial x}. 
\]%
After the redefinition $C_{2}^{n}=C_{2n}=C_{p}$ we can obtain the standard
form of the hydrodynamic chain: 
\begin{equation}
\{C_{p}(x),C_{q}(y)\}=\frac{pq(p-1)}{p+q-2}C_{p+q-2}(x)\frac{\partial \delta
(x-y)}{\partial x}-\frac{pq(q-1)}{p+q-2}C_{p+q-2}(y)\frac{\partial \delta
(y-x)}{\partial y}.  \label{eq42}
\end{equation}%
The second example of the infinite dimensional chain is based on the
invariant chiral currents of the WZNW model with the $SU(\infty )$, $%
SO(\infty )$, $SP(\infty )$ constant torsions. If dimension of matrix
representation $n$ is not ended $(n\rightarrow \infty )$ , all the chiral
currents are the primitive currents. This is easy to see from the expression
for the new chiral currents $C_{m,k}$ (see e.g. [13, 14]). The PB in
Liouville coordinates $C_{m}(x),\;m=2,3,...,\infty $ takes the form: 
\begin{equation}
\{C_{m}(x),C_{n}(y)\}=-W_{mn}(C(y))\frac{\partial }{\partial y}\delta
(y-x)+W_{nm}(C(x))\frac{\partial }{\partial x}\delta (x-y),  \label{eq43}
\end{equation}%
\begin{equation}
W_{mn}(C(x))=\frac{mn(n-1)}{m+n-2}C_{m+n-2}(x).  \label{eq44}
\end{equation}%
This PB obey the skew-symmetric condition: $\{C_{m}(x),C_{n}(y)\}=-%
\{C_{n}(y),C_{m}(x)\}$. However, the Jacobi identity imposes conditions on
the Hamiltonian function $W_{mn}(C(x))$ [18]: 
\begin{equation}
(W_{kp}+W_{pk})\frac{\partial W_{mn}}{\partial C_{k}}=(W_{km}+W_{mk})\frac{%
\partial W_{pn}}{\partial C_{k}},\;\frac{dW_{kp}}{dx}\frac{\partial W_{nm}}{%
\partial C_{k}}=\frac{dW_{km}}{dx}\frac{\partial W_{np}}{\partial C_{k}}.
\label{eq45}
\end{equation}%
The Jacobi identity is satisfied by the metric tensor $W_{mn}(C(x))$ (\ref%
{eq44}). The algebra of charges $\int\limits_{0}^{2\pi }C_{n}(x)dx$ is the
abelian algebra. Now, let us choose the Casimir operators $C_{n}$ as the
Hamiltonians: 
\begin{equation}
H_{n}=\frac{1}{n}\int\limits_{0}^{2\pi }C_{n}(x)dx,\;n=2,3...\,.
\label{eq46}
\end{equation}%
Then, the equations of motion for the densities of Casimir operators are the
following: 
\begin{equation}
\frac{\partial C_{m}(x)}{\partial t_{n}}=\frac{m(n-1)}{m+n-2}\frac{\partial
C_{m+n-2}}{\partial x}.  \label{eq47}
\end{equation}%
Thus, the invariant chiral currents with the $SU(2)$, $SO(3)$, $SP(2)$
constant torsion and the invariant chiral currents with the $SU(\infty )$, $%
SO(\infty )$, $SP(\infty )$ constant torsion form the same infinite
hydrodynamic chain (\ref{eq42}), (\ref{eq43}), ({\ref{eq44}). This PB (\ref%
{eq43}) is particular case of the $M$-bracket given by Dorfman [19] and
Kupershmidt [20] for $M=2$ and describe the hydrodynamic chains. We can
construct new nonlinear equations of motion for the initial chiral currents $%
U^{\mu }$ using the flat $PB_{2}$ (\ref{eq21}) and the Hamiltonians $H_{n}$ (%
\ref{eq46}), where $C_{n}(x)$ is defined by the equation (\ref{eq26}) for
the $SU(\infty )$ group: 
\[
\frac{\partial U_{\mu }(x)}{\partial t_{n}}=\frac{1}{n}\int\limits_{0}^{2\pi
}dy\{U_{\mu }(x),C_{n}(U(y))\}_{2}, 
\]%
\begin{equation}
\frac{\partial U_{\mu }(x)}{\partial t_{n}}=\frac{\partial }{\partial x}%
[d_{\nu _{1}\nu _{2}}^{k_{1}}d_{k_{1}\nu _{3}}^{k_{2}}...d_{\nu _{n-1}\mu
}^{k_{n-3}}U^{\nu _{1}}(x)...U^{\nu _{n-1}}(x)].  \label{eq51}
\end{equation}%
As an example we consider $n=3$: 
\begin{equation}
\frac{\partial U_{\mu }}{\partial t_{3}}=\frac{\partial }{\partial x}(d_{\mu
\nu \lambda }U^{\nu }U^{\lambda }),\;\;\mu =1,2,...\infty .  \label{eq50}
\end{equation}%
It is easy to see that this dynamical system is a bi-Hamiltonian one: 
\begin{equation}
\frac{\partial U_{\mu }(x)}{\partial t_{3}}=\frac{1}{3}\int\limits_{0}^{2\pi
}dy\{U_{\mu }(x),C_{3}(U(y))\}_{2}=\frac{1}{2}\int\limits_{0}^{2\pi
}dy\{U_{\mu }(x),C_{2}(U(y))\}_{3}.  \label{eq51}
\end{equation}%
Above the $PB_{3}$ has form: 
\begin{equation}
\{U_{\mu }(x),U_{\nu }(y)\}_{3}=2d_{\mu \nu \lambda }U^{\lambda }.
\label{eq52}
\end{equation}%
Let us remind that $d_{\mu \nu \lambda }$ are the symmetric structure
constant of the $SU(\infty )$ algebra in a matrix representation. This PB
satisfies to Jacobi identity for ($n\rightarrow \infty $): 
\[
d_{\sigma \mu \nu }d_{\sigma \lambda \rho }+d_{\sigma \mu \lambda }d_{\sigma
\nu \rho }+d_{\sigma \mu \rho }d_{\sigma \nu \lambda }=\frac{1}{n}(\delta
_{\mu \nu }\delta _{\lambda \rho }+\delta _{\mu \lambda }\delta _{\nu \rho
}+\delta _{\nu \rho }\delta _{\nu \lambda }). 
\]%
Analogically we can obtain the equation of motion for the chiral currents of 
}${SO(\infty )}${\ and }$SP(\infty )$:{%
\begin{equation}
\frac{\partial U_{\mu }(x)}{\partial t_{n}}=\frac{\partial }{\partial x}%
[v_{\nu _{1}\nu _{2}\nu _{3}}^{k_{1}}...v_{\nu _{2n-2}\nu _{2n-1}\mu
}^{k_{2n-3}}U^{\nu _{1}}...U^{\nu _{2n-1}}].  \label{eq53}
\end{equation}%
To see how it works, for example, let us consider $n=4$: 
\begin{equation}
\frac{\partial U_{\mu }}{\partial t_{4}}=\frac{\partial }{\partial x}(v_{\mu
\nu \lambda \rho }U^{\nu }U^{\lambda }U^{\rho }),\;\mu =1,2,...\infty .
\label{eq54}
\end{equation}%
Also we can obtaine a solution for the metric function $W_{mn}(C(x))$ which
is analog to the Dubrovin-Novikov metric tensor $W_{\mu \nu }=\frac{\partial
^{2}F}{\partial U^{\mu }\partial U^{\nu }})$ : 
\[
C_{m}(U(x))=mF((U(x)),\;\;F(x,t_{n})=g(t_{n}+\frac{x}{n-1}) 
\]%
and $g(t_{n}+\frac{x}{n-1})$ is an arbitrary function of its argument. }
\noindent

\begin{center}
\textbf{Acknowledgements}
\end{center}

\noindent Gershun V.D. should like to thank B.A. Dubrovin, O.I. Mokhov, M.V.
Pavlov, S.P. Tsarev,A.A. Zheltukhin for interest in his investigation and
fruitful discussions. Cirilo-Lombardo D.J. is very grateful to the JINR
Directorate and the BLTP for his hospitality and finnancial support.

\begin{center}
{\footnotesize {REFERENCES} }
\end{center}

{\footnotesize {\ 1. Wess J., Zumino B. Lagrangian method for chiral
symmetries //{Phys. Rev.}, 1967, Vol. 163, p. 1727-1735.\newline
2. Coleman S., Wess J., Zumino B. Structure of phenomenological lagrangians
1. //{Phys. Rev.}, 1969, Vol. 177, p. 2239-2247.\newline
3. Volkov D.V. Phenomenological lagrangian interection of goldstone
particles //{preprint/ NAC Ukraine, Kiev, ITP}, 1969, 51 pp. (in Russian).%
\newline
4. Volkov D.V., Gershun V.D., Tkach V.I. Current structure of
phenomenological lagrangians //{Theoretical and Mathematical Physics}, 1970,
Vol. 3, p.321-328. (in Russian).\newline
5. Wess J., Zumino B.Consequences of anomalous Ward identities //{Phys.
Lett. B}, 1971, Vol. 37(1), p. 95--97.\newline
6. Witten E. Global aspects of current algebra //{Nucl. Phys. B}, 1983, Vol.
223(2), p. 422--432.\newline
7. Witten E. Non-abelian bosonization in two dimensions //{Commun. in Math.
Phys.}, 1984, Vol. 92(4), p. 455--472.\newline
8. Novikov S.P. Multivalued functions and functionals. An analoque of the
Morse theory. //{Sov. Math. Dokl.}, 1981, Vol. 24, p. 222--226.\newline
9. Kac V. Simple graded Lie algebras of finite growth //{Funkt. Anali. i ego
Prilozhen.}, 1967, Vol. 1, p. 82-126.(English translation: //{Functional
Anal. Appl.} 1967, Vol. 1, p. 328-372.)\newline
10. Moody R.V. Lie algebras associated with generalized Cartan matrices //{%
Bull. Amer, Math, Soc.}, 1967, Vol. 73, p. 217-221.\newline
11. Zakharov V.E., Mikhailov A.V. Relativistically invariant two-dimensional
models in field theory integrable by inverse problem technique //{JETP},
1974, Vol. 74, p. 1953-1973. (in Russian)\newline
12. Volkov D.V., Gershun V.D., Tkach V.I. About nonlocal charge algebra in
two-dimensional models //{Ukrainian Journal of Physics}, 1983, Vol. 28, p.
641-649. (in Russian)\newline
13. Gershun V.D. Integrable string models in terms of chiral invariants of $%
SU(n)$, $SO(n)$, $SP(n)$ groups //{SIGMA}, 2008, Vol. 4, 16 pp.\newline
14. Gershun V.D. Integrable string models of WZNW type with constant $SU(2)$%
, $SO(3)$, $SP(N)$ and $SU(3)$ torsions and hydrodynamic chains// {Phys. of
Part. and Nucl.,2012, Vol. 43, p. 659-662.\newline
15. Goldschmidt Y.Y. and Witten E. Conservation laws in some two-dimensional
models //{Phys. Lett. B}, 1980, Vol. 91, p. 392--396.\newline
16. Evans J.M., Hassan M., MacKay N.J., Mountain A.J. Local conserved
charges in principal chiral models //{Nucl. Phys. B}, 1999, Vol. 561, p.
385--412.\newline
17. de Azcarraga J.A., Macfarlane A.J., Mountain A.J. and Perez Bueno J,C..
Invariant tensors for simple groups //{Nucl. Phys. B}, 1998, Vol. 510, P.
657-687.\newline
18. Dubrovin B., Zhang Y. Bihamiltonian hierarhies in 2D tological field
theory at one-loop approximation// {Commun. Math. Phys.,1998, Vol. 198, p.
311-361.\newline
19. Dorfman I.Ya. Dirac structures and integrability of nonlinear evolution
equations// {Nonlinear sciences: Theory and Applications}, John Wiley and
Sons, New York, 1993,176 pp.\newline
20. Kupershmidt B.A,, Manin Yu.I. Long wave equations with a free surface
II. The hamiltonian structure and the higher equations// {Func. Anal.
Appl.,1978, Vol. 12, p. 25-37. }}}}}

\end{document}